\documentclass[12pt]{article}
\input{epsf.tex}
\begin{document}
\begin{center}
{\bf Estimation of the value and localization of possible systematic errors in
 determination of level density and radiative strength functions from the
 $(n,2\gamma )$-reaction.}
\end{center}
\begin{center}
{\bf V.~A. Khitrov, Li Chol,  A.~M. Sukhovoj}\\
{\it Frank Laboratory of Neutron Physics, Joint Institute for Nuclear Research,
141980, Dubna, Russia}
\end{center}
\begin{abstract}
 Systematic error in determination of the absolute intensities of the
 two-step $\gamma$-cascades after the thermal neutron capture and its influence
 on the value and localization of extracted from
 $(n,2\gamma )$-reaction probable level densities and radiative strength
 functions of dipole $\gamma$-transitions have been analysed. It was found that
 this error in limits of its possible magnitude cannot change made earlier
 conclusions about the radiative strength functions of $E1$ and $M1$
 transitions at $E_{\gamma}\simeq 3$ MeV and level density of heavy nucleus below
 $\simeq 0.5B_n$.
\end{abstract}
 \section{Introduction}
  Both radiative strength functions $k$ of $E1$ and $M1$ transitions and density
  $\rho$ of levels populated by them are determined by inner nuclear processes
  at corresponding excitation energy. Below the neutron binding energy $B_n$,
  the parameters $k$ and $\rho$ are, in practice, the only probe for nuclear
  models. Therefore, determination of these parameters with the maximum
  possible reliability is one of the principle tasks of low-energy
  nuclear physics.

  Very high level density of deformed, first of all, nuclei and a number of
  quanta following $\gamma$-decay do not allow one to determine mentioned
  parameters by the direct methods of nuclear spectroscopy above
  $E_{ex}\simeq 2-3$ MeV. The only possibility to solve this task is the
  selection of probable values of $k$ and $\rho$ which allow one to
  reproduce experimental spectra within the experimental errors and
 uncertainties of the used model notions. The main source of systematic
 errors of such procedure is the strong correlation between $k$ and $\rho$.
 This results from the fact that the intensity of the primary production
 of the reaction is determined by a product of the probability of emission
 of reaction product (including $\gamma$-quanta) and $\rho$.

 This causes a necessity to develop new experimental methods in which the
 observables are connected by the mathematics relations of other type.
 The method [1] developed at the FLNP JINR for simultaneous model-free
 estimation of the  radiative strength functions $k$ and density of levels
 $\rho$ by means of  intensities $I_{\gamma\gamma}$ of populating them
 cascades which follow thermal neutron capture. These intensities are
 determined by  energy dependence of $k$ for $E1$ and $M1$ transitions and,
 qualitatively, are inversely proportional to $\rho$. As a consequence,
 the interval of $k$ and  $\rho$ values allowing reproduction of the
 experimental value of  $I_{\gamma\gamma}$ is narrow. This can promote
 identification of false values  of $k$ and $\rho$ determined [2] from the
 nucleon evaporation spectra or spectra of primary $\gamma$-transitions [3].
 Of course, the data obtained according to [1] contain some statistic
 uncertainty caused by inadequacy of the model ideas of of the $\gamma$-decay
 process to the experiment and error in the experimental data used for
 determination of $I_{\gamma\gamma}$.

\section{Conditions of the experiment}
\subsection{Model approaches}

 The sum intensities of the cascades whose intermediate levels are lying in
 the excitation energy interval $\Delta E$ are calculated under assumptions:

 (a)~the branching ratios for the decaying level $i$ do not depend on the mode
 of its population;

 (b)~at a given excitation energy, only the levels belonging to solely
 statistical ensemble exist. I.e., the mean reduced probability of their
 population by the primary $E1$ or $M1$ transitions is equal for any level in
 the spin window set by the selection rule and does not depend on the
 structure of the wave function the neutron resonance. On the basis of the
 mathematics theorem of the mean, any sum of partial  widths is represented
 in calculation (like that used in [1]) by the product  of  their number and
 some mean partial width (which is determined through corresponding strength
 function);

 (c)~the energy dependence of $k$ (but not its absolute value) is equal for
 the primary and secondary cascade transitions.

 One can suppose that the maximal discrepancy between the model notions and
 experiment is conditioned by assumptions  (b) and (c). The truth of the
 assumption (c) now can be tested in direct experiment, but possibility to
 test thesis (b) is quite hypothetical.

 In particular, the $k$ and $\rho$ values obtained according to [1] allow
  good reproduction of intensity distribution of cascades to the final levels
  with $E_{ex}\leq 1$ MeV and total experimental absolute intensities to the
  final levels with $E_{ex}\geq 2$ MeV  excitation energy (without additional
   fit) in the $^{191,193}Os$ and $^{118}Sn$ nuclei. Somewhat different
   situation is observed in $^{185,187}W$ where analogous data allow
   possibility of some violation of condition (c).

   Potentially, possibility of violation of condition (b) follows from
   qualitative interpretation of the results obtained in [1]. Change in the
   power of energy dependence in vicinity of $0.5B_n$ allows an assumption
   about the corresponding change in properties of levels excited by
   cascades in this region. If this change is spread over the excitation
   energy then violation of condition (b) in the vicinity of $0.5B_n$ can
   be significant. Although it should be noted that the approximation [4]
   of the experimental cumulative sums of cascade intensities shows that
   this distribution above $i_{\gamma\gamma}\simeq 10^{-4}$ is well reproduced
   by the sum of only two distributions which correspond to primary $E1$ and
    $M1$ cascade transitions, respectively.

\subsection{Errors of the experiment}

  In modern experiment [5] the error of cascade intensity
 $\delta I_{\gamma\gamma}$ it is mainly determined by the error of known [6]
 intensities of the high-energy primary transitions $i_1$ used for
 normalization of the $I_{\gamma\gamma}$ value.
 As it was shown in [5], uncertainty in determination [7] of dependence  of
  $I_{\gamma\gamma}$ on the primary transition energy from the experimental
  spectra (each of them is the sum of two distributions) can be negligible
  if $HPGe$ detectors with 20-25\% and higher efficiency are used in the
  experiment. This is true at least for even-odd deformed and even-even
  spherical nuclei. This conclusion unambiguously follows from the
  extrapolation [4] of intensity distribution of cascades (resolved in the
   spectrum as pairs of peaks) to zero value $i_{\gamma\gamma}=0$ ---
   more than 95-99\% of intensity of cascades with intermediate levels lying
   below $0.5B_n$ is determined in the experiment. So, at any energy of the
   cascade primary transition the shape of the functional dependence

 \begin{equation}
 I_{\gamma\gamma}=\sum_{\lambda ,f}\sum_{i}\frac{\Gamma_{\lambda i}}
 {\Gamma_{\lambda}}\frac{\Gamma_{if}}{\Gamma_i}=\sum_{\lambda ,f}
 \frac{\Gamma_{\lambda i}}{<\Gamma_{\lambda i}>
 m_{\lambda i}} n_{\lambda i}\frac{\Gamma_{if}}{<\Gamma_{if}> m_{if}}
\end{equation}
 is ascertained with relative error not more than several percent.
 Therefore, systematic error in determination of $k$ and $\rho$ from
 combination of $\Gamma_{\lambda}$ and $I_{\gamma\gamma}(E_1)$  (1) is
 practically caused by error in obtaining absolute $I_{\gamma\gamma}$ values,
 i.e., by  error of $i_1$. Comparison [8] between the known [6] and measured
  in Budapest values of $i_1$ shows that the uncertainty of any arbitrary
  taken intensity $i_1$ for the data [6] can be estimated as 20\%.
   Normalization of $I_{\gamma\gamma}$ to absolute value is made using 5-10
 (supposedly non-correlated in errors) values of $i_1$. So, one can expect,
 that the total  experimental intensity $I_{\gamma\gamma}$ differs from the
 actual value not more than by 40\%, if $I_{\gamma\gamma}$ is normalized
  using data [6].

  \section{Uncertainties in determination of $k$ and $\rho$ and their
  localization}
 The values $k$ and $\rho$ resulting in a given magnitude of
 $\delta I_{\gamma\gamma}$, in practice, cannot be estimated using
 conventional in mathematical statistics procedure of error transfer.
 Therefore, these parameters were estimated in other manner. Varying the
 total experimental intensities of cascades from 70 to 120\% in $^{185,187}W$
 and  from 50 to 100\% in $^{191,193}Os$ we obtained an ensemble of functional
 dependences of $k$ and $\rho$ which permit reproduction of these
 $I_{\gamma\gamma}$ values. Variation coefficient $\kappa$  is stipulated by
 the fact that  the total intensities of the observed primary transitions and
  two-step cascades  to low-lying levels ($E_f<0.5$ MeV) for isotopes of $W$
  and $Os$  equal 65-70\% and 92-95\%, respectively. Besides, statistics of
  the accumulated  coincidences and high energy resolution permit one to
  neglect the errors of  procedure of decomposition of the experimental
  spectra into two components  corresponding to solely primary and solely
  secondary cascade transitions.

 Probable $\rho$ and $k$ values obtained in this manner are shown in Figs.~1-4
 and 5-8, respectively. These data  are compared in Figs.~9 and 10 with the
 model values of $\rho$ and $k$ in function of coefficient $\kappa$ for four
 excitation energies and four energies of $\gamma$-quantum.
  The main result of this analysis -- $\rho$ in this interval of variation
  of $\kappa$ (probably overlapping the region of possible error of
  $I_{\gamma\gamma}$)  weakly changes at the excitation energy below
  $\simeq 0.5B_n$.  Variations of the energy dependence of $k$ in this case are
  much  stronger. In the all considered cases, however, the sum of $k(E1)+k(M1)$
   for $E_1<2-3$  MeV does not exceed the sum of the calculated according to [10]
   value $k(E1)$  and value $k(M1)=const$ if their ratio is normalized to
   the experimental (approximated) value. Taking into account that the sign
   of derivative of  $k^{exp}/k^{mod}$ in Fig.~10 differs for different
   energies $E_{\gamma}$ and its value is close to zero at 2.5-3.0 MeV for
   three nuclei discussed here and some bigger for $^{191}Os$ one can expect
   that the strength functions is estimeted with the maximum possible
   reliability independently on $\delta I_{\gamma\gamma}$ at least at this
   $\gamma$-transition energy.

\section{Discussion}
 In the framework of the enumerated above assumptions about the cascade
 $\gamma$-decay process one can conclude that significant decrease in the level
 density of heavy, at least, nuclei as compared with the predictions of the
 Fermi-gas model cannot be explained by some revealed up to now uncertainties
 of the method [1].

 The observed results [1] on the level density can be qualitatively explained
 within modern nuclear models directly accounting [13] for dynamics of
 breaking of the nucleon pairs or within the generalized model of superfluid
 nucleus [2]. In both cases the corrections of no principle  of these models
 are needed. In the former, the second pair must break at $\simeq 1$ MeV
 higher excitation energy as compared with [13]. In the latter, the temperature
 of the expected phase transition of a nucleus from superfluid to usual state
 should be decreased by a factor of about 1.5. Besides, the notions of the
 calculation of entropy below the point of the expected phase transition should
 be slightly corrected. This provides appearance of the clearly expressed
 step-like structure in energy dependence of level density that is observed in
 experiment [1]. Both conditions have analogies in the experimental data on
 super-conductivity (including high-temperature one) and super-fluidity of
 usual matter and, therefore, can be considered as quite possible.

 Of course, one cannot exclude existence of unknown factors which distort
 the simple enough model [1] of the $\gamma$-decay process. First of all,
 it is necessary to test in direct experiment the possible and probably strong
 influence of the wave function structure of neutron resonance on the
 $\gamma$-decay process of high-lying states of a nucleus. Observation of
 strong correlation of reduced neutron width  of resonance $\Gamma^0_n$ and
 partial widths of the cascade primary transitions to the levels with
 $E_{ex}>1-3$ MeV would decrease discrepancy between the results [1] and notions
 of old models from one hand but require significant modification of the model
 ideas of a nucleus from other hand. These problems can be solved in
 experimental study of two-step cascades in different resonance of the same
 nucleus. This requires, however, to increase statistics of $\gamma -\gamma$
 coincidences. Development of new of principle methods to study the
 $\gamma$-decay process can bring to the analogous result.\\
1. E.V. Vasilieva, A.M. Sukhovoj, V.A. Khitrov,
Phys. At. Nuc. (2001) {\bf 64(2)} 153\\
2. M.I. Svirin, G.N. Smirenkin,  Yad. Fiz. (1988)  {\bf47} 84\\
3. G.A. Bartholomew et al.,
 Advances in nuclear physics (1973) {\bf 7} 229\\
4. A.M. Sukhovoj, V.A. Khitrov, Phys. of Atomic Nuclei, (1999) {\bf 62(1)} 19\\
5. V.A. Khitrov, A.M. Sukhovoj, JINR preprint E3-2002-276, Dubna, 2002.\\
6. M.A. Lone et al., Nucl. Data Tables. (1981) {\bf 26(6)} 511\\
7. S.T. Boneva et. al., Nucl.  Phys. (1995) {\bf A589} 293\\
8. G.L. Molnar et al., App. Rad. Isot. (2000) {\bf 53} 527\\
9. W. Dilg, W. Schantl, H. Vonach and M. Uhl, Nucl. Phys., (1973) {\bf A217}
 269\\
10. S.G. Kadmenskij, V.P. Markushev and W.I. Furman, Sov. J. Nucl Phys. Yad., (1983)  {\bf 37} 165\\
11. P. Axel,  Phys. Rev. (1962) {\bf 126(2)} 671\\
12. J. M. Blatt and V. F. Weisskopf,
 Theoretical Nuclear Physics, New York (1952)\\
13. A.V. Ignatyuk, Yu.V. Sokolov, Yad. Fiz., (1974)  {\bf 19} 1229
\\\\

\begin{figure}
\begin{center}
\leavevmode
\epsfxsize=13.5cm
\epsfbox{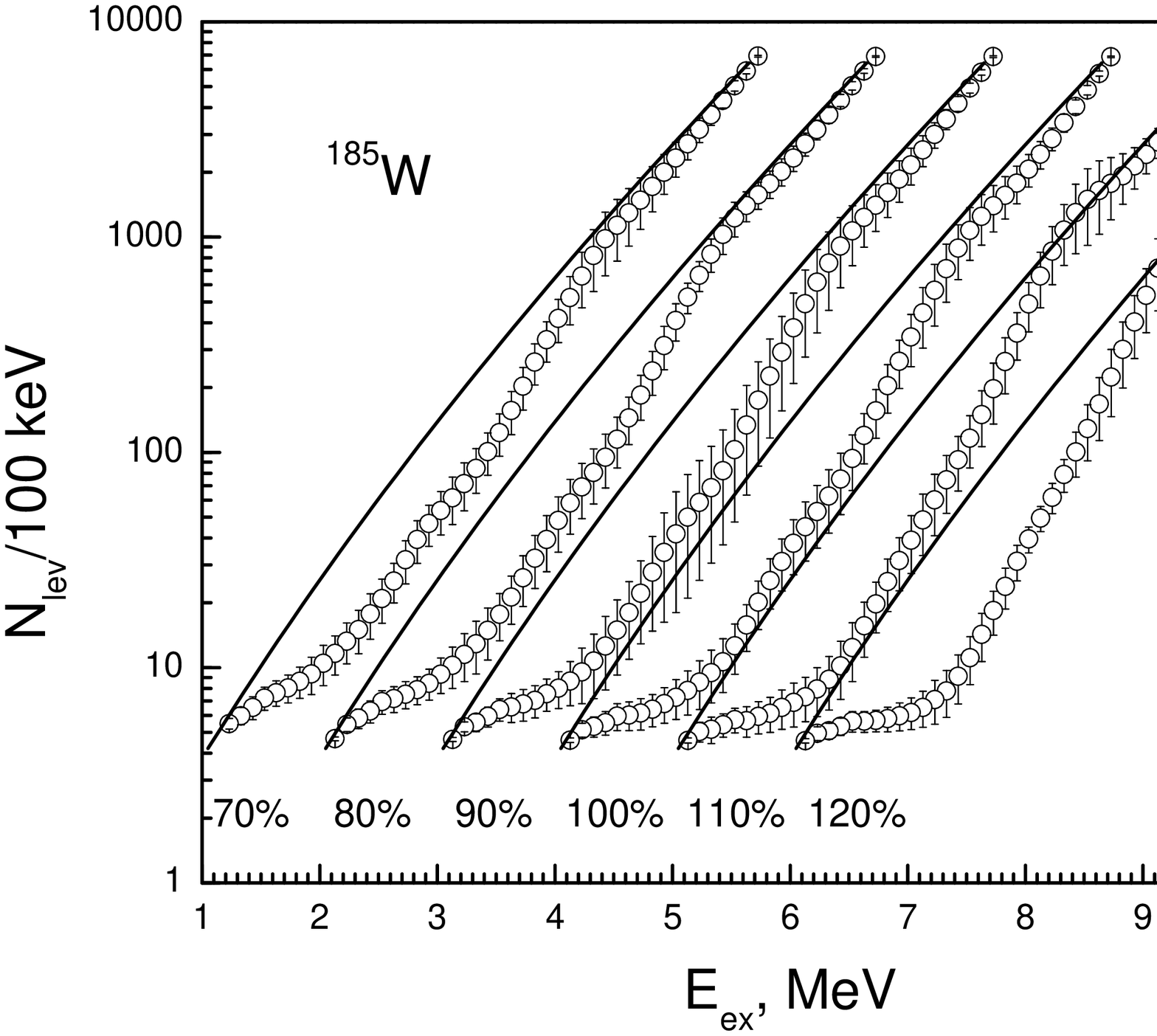}
\end{center}
\vspace{-3.5cm}
Fig.~1.~The mean value and interval of variations of level density
 $(1/2^{\pm}\leq J^{\pi}\leq 3/2^{\pm})$ populated by two-step cascades in
 $^{185}W$ (points with bars). Line represents calculation according model [9].
 Each variant of calculation is shifted relative to previous one by 1 MeV
 and is marked by value of $\kappa$ (\%).
\end{figure}

\begin{figure}
\vspace{-2cm}
\begin{center}
\leavevmode
\epsfxsize=13.5cm
\epsfbox{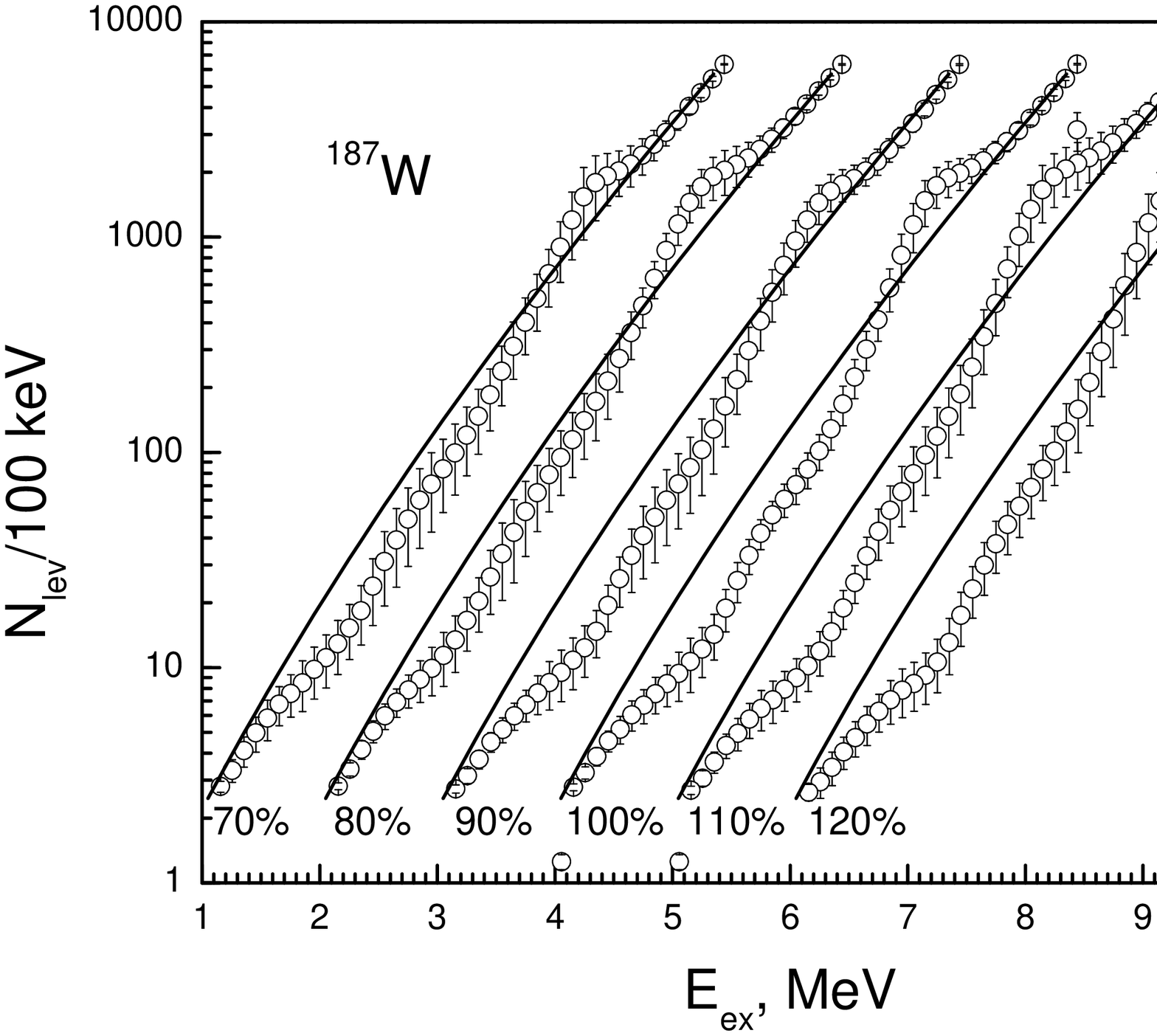}
\end{center}
\vspace{-3.5cm}
 Fig.~2.~The same as in Fig.~1 for $^{187}W$.
\end{figure}

\newpage

\begin{figure}\vspace{3cm}
\begin{center}
\leavevmode
\epsfxsize=13.5cm
\epsfbox{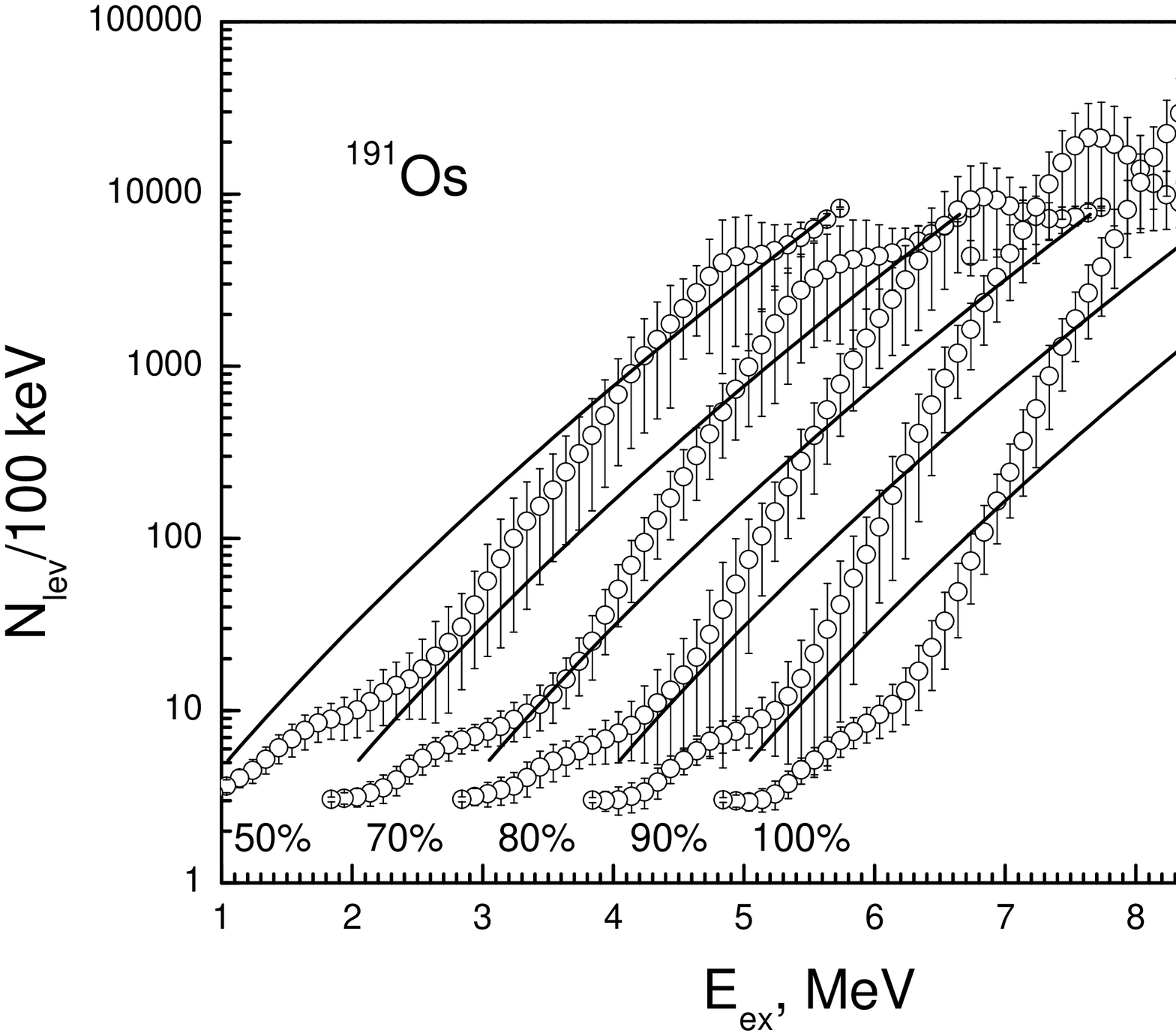}
\end{center}
\vspace{-3.5cm}
 Fig.~3.~The same as in Fig.~1 for $^{191}Os$.
\end{figure}

\vspace{-2cm}
\begin{figure}
\begin{center}
\leavevmode
\epsfxsize=13.5cm
 \epsfbox{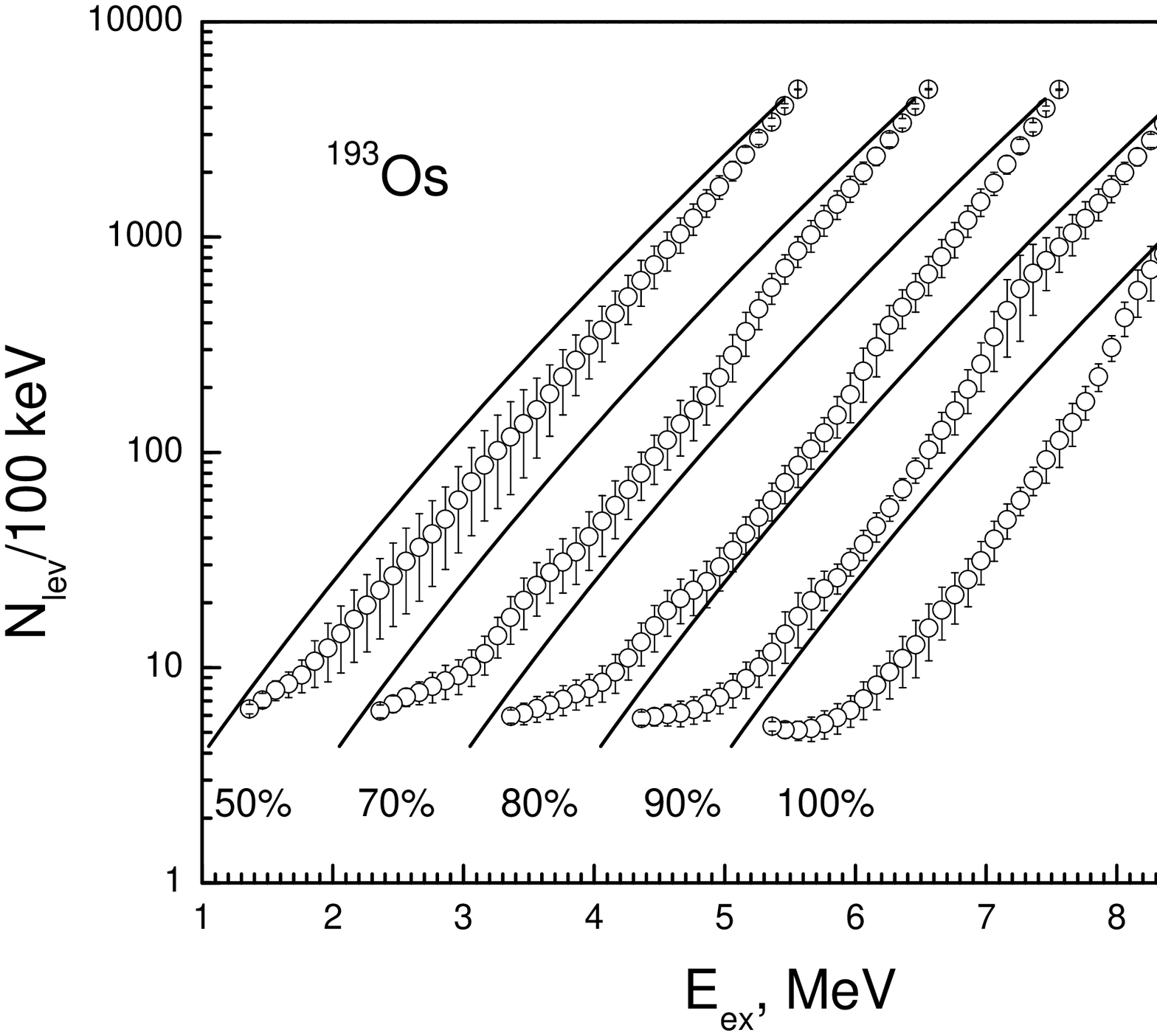}
\end{center}
\vspace{-3.5cm}
 Fig.~4.~The same as in Fig.~1 for $^{193}Os$.
\end{figure}

\newpage

\begin{figure}\vspace{3cm}
\begin{center}
\leavevmode
\epsfxsize=13.5cm
\epsfbox{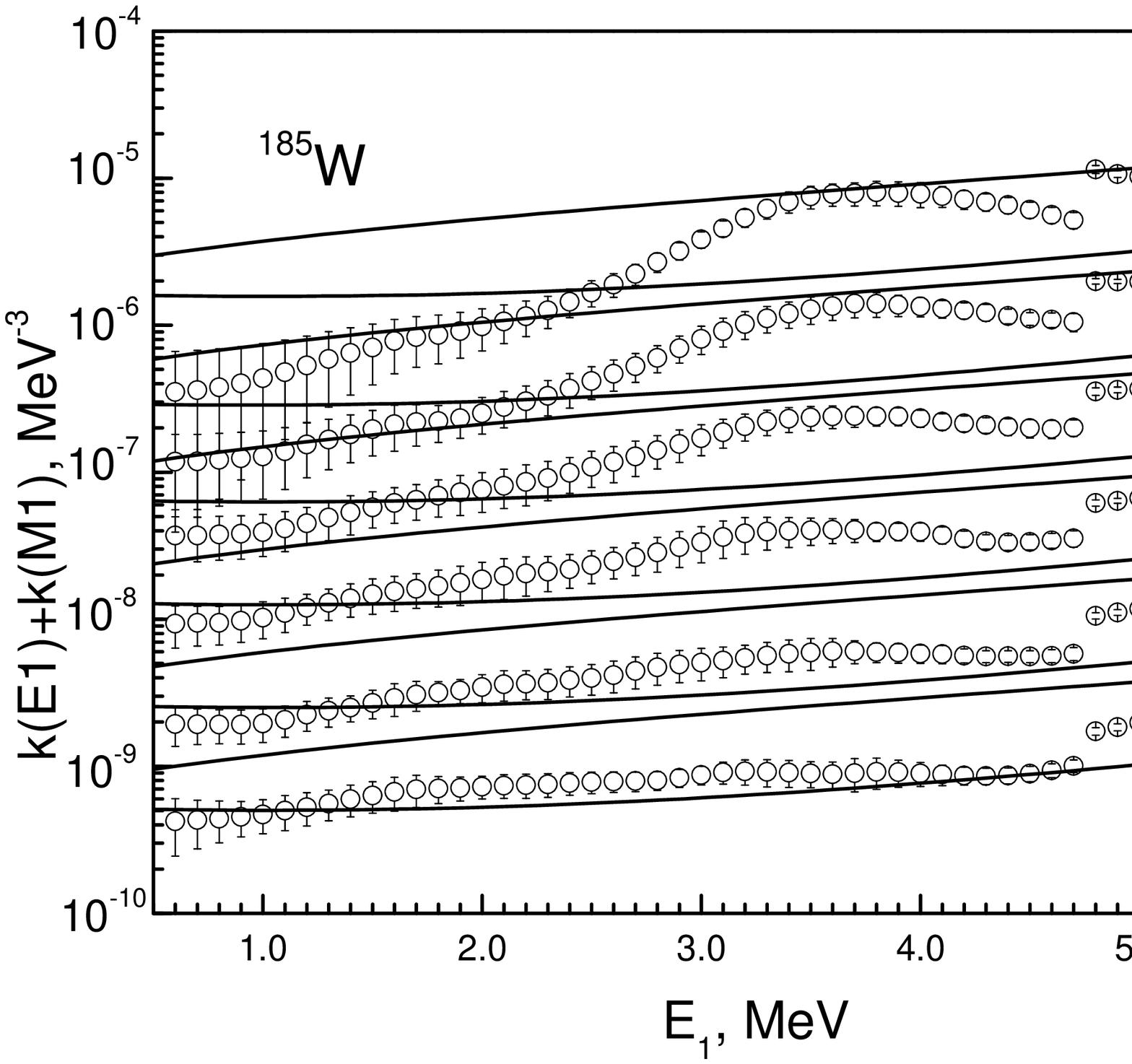}
\end{center}
\vspace{-3.5cm}
 Fig.~5.~ The mean value and interval of variations of radiative strength
 functions for $^{185}W$ (points with bars). Predictions models [10] and
 [11] in sum with $k(M1)=const$ [12]. Experimental and model values of each
 variant are increased by a factor of 5.
\end{figure}

\begin{figure}\vspace{3cm}
\begin{center}
\leavevmode
\epsfxsize=13.5cm
\epsfbox{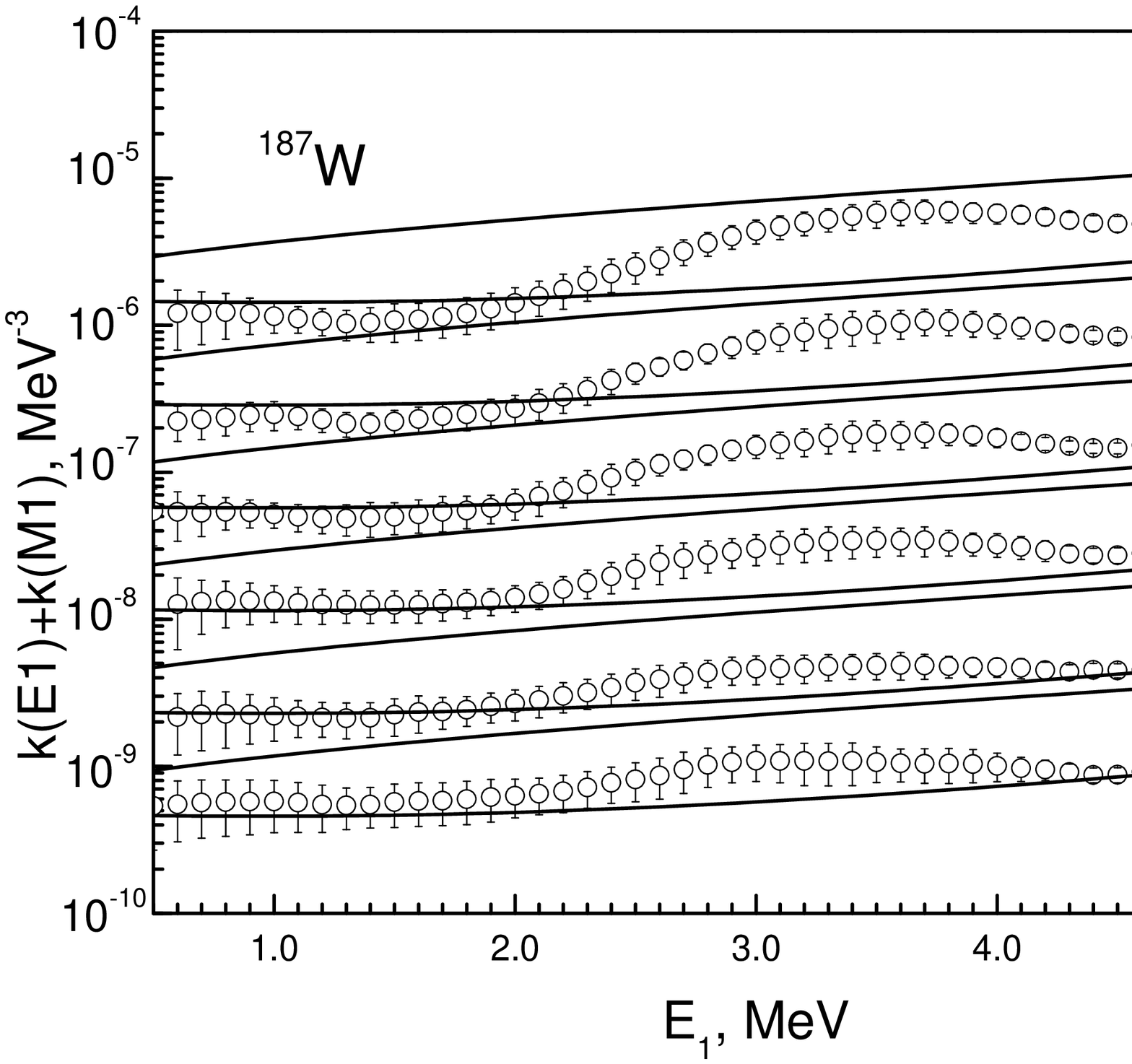}
\end{center}
\vspace{-3.5cm}
 Fig.~6.~The same as in Fig.~5 for $^{187}W$.
\end{figure}

\newpage

\begin{figure}\vspace{3cm}
\begin{center}
\leavevmode
\epsfxsize=13.5cm
\epsfbox{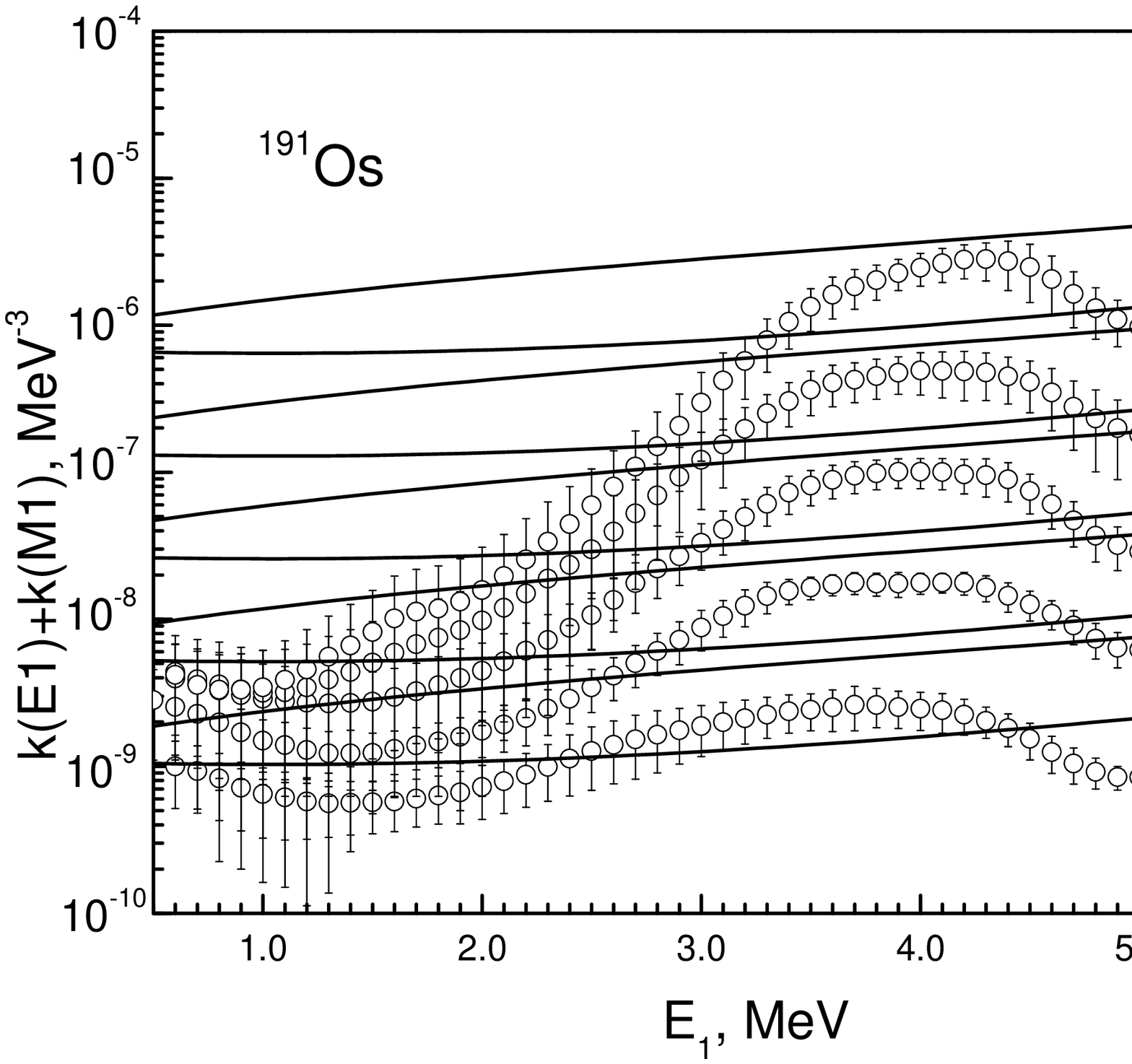}
\end{center}
\vspace{-3.5cm}
 Fig.~7.~The same as in Fig.~5 for $^{191}Os$.
\end{figure}

\begin{figure}\vspace{3cm}
\begin{center}
\leavevmode
\epsfxsize=13.5cm
\epsfbox{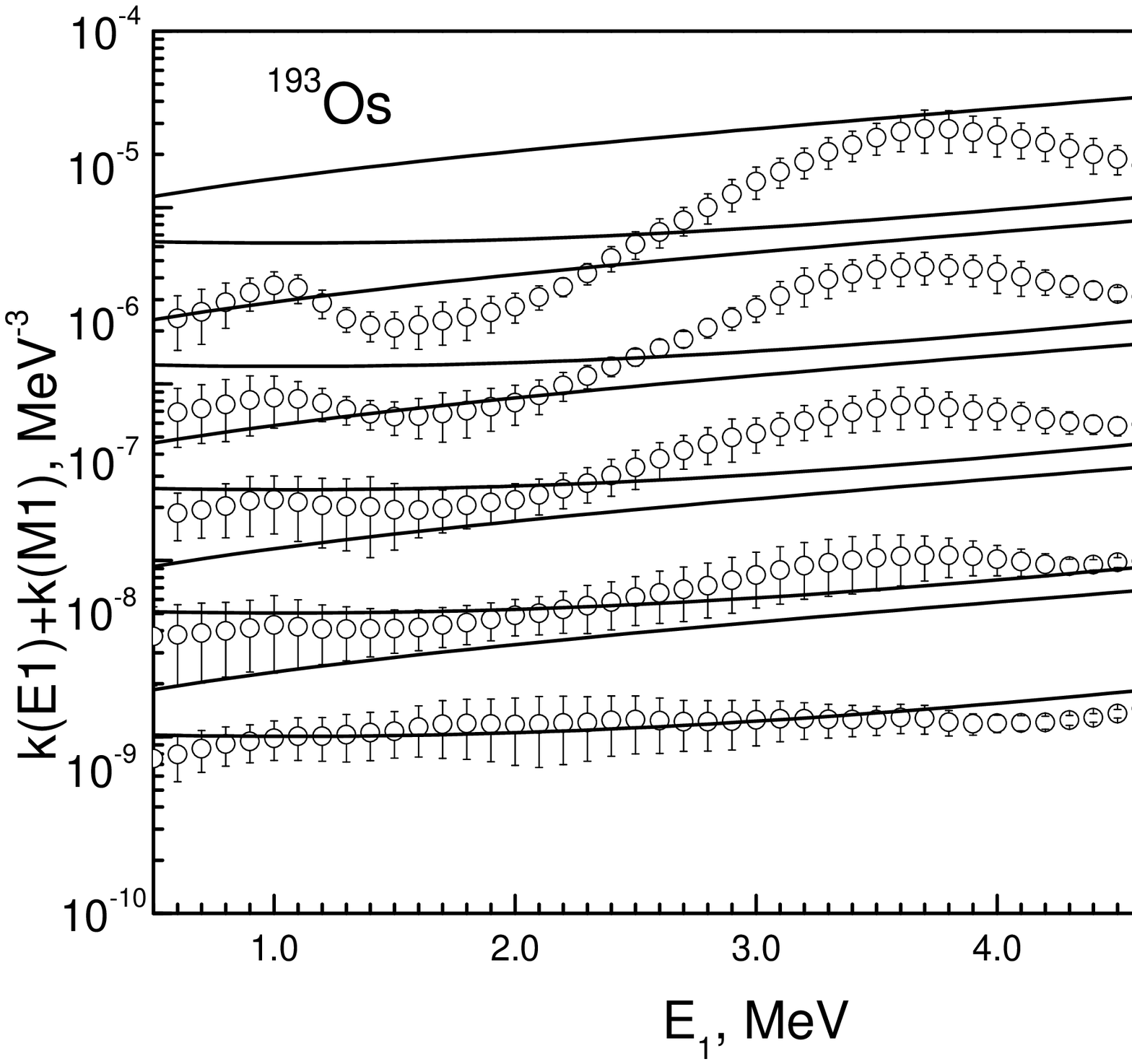}
\end{center}
\vspace{-3.5cm}
 Fig.~8.~The same as in Fig.~5 for $^{193}Os$.
\end{figure}
\newpage

\begin{figure}\vspace{2cm}
\begin{center}
\leavevmode
\epsfxsize=14.5cm
\epsfbox{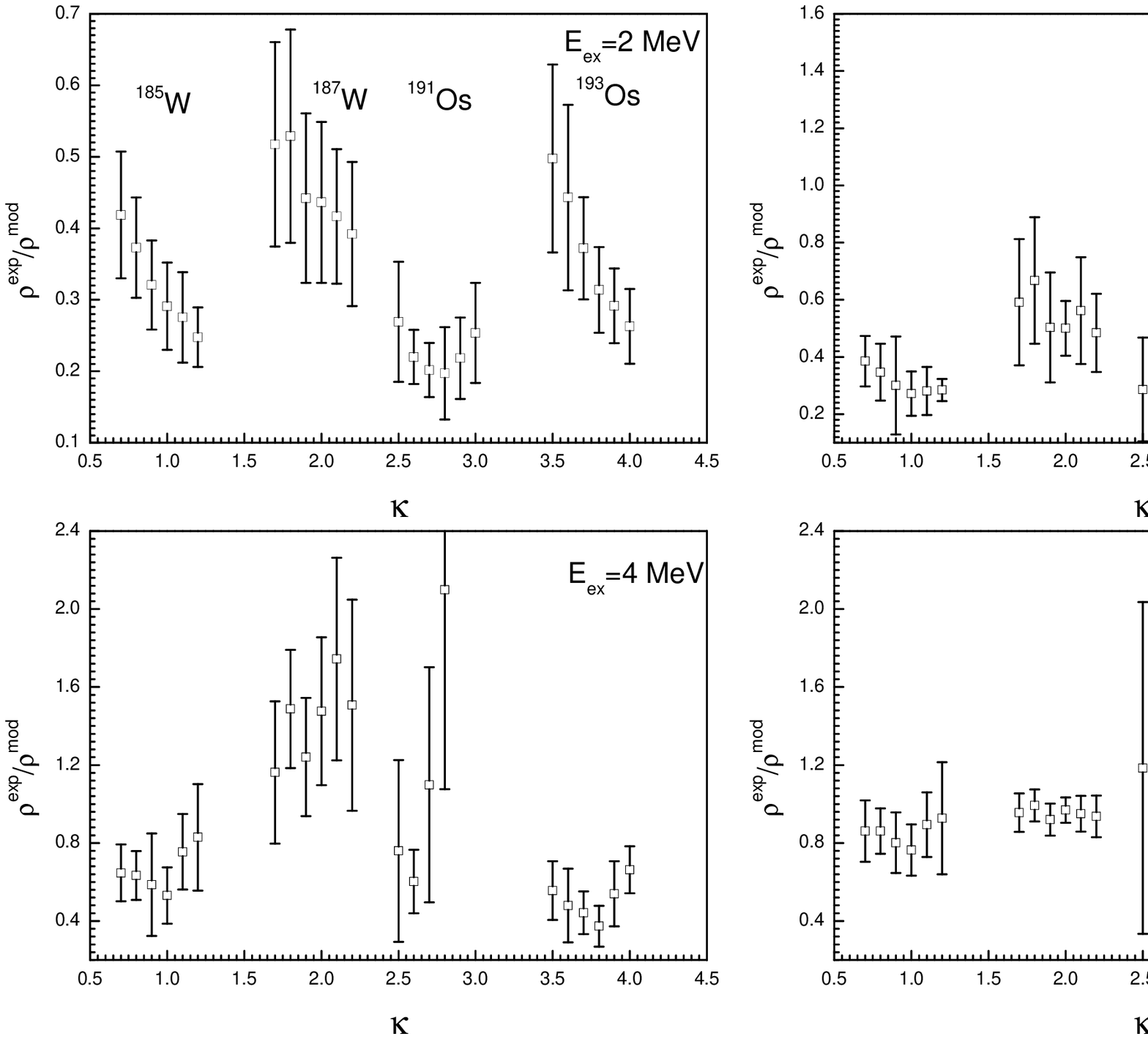}
\end{center}\vspace{-5cm}
 Fig.~9.~The ratio between the experimental and model level densities in
 function of parameter $\kappa$ for 4 excitation energies of studied nuclei.
\end{figure}

\begin{figure}\vspace{2cm}
\begin{center}
\leavevmode
\epsfxsize=14.5cm
\epsfbox{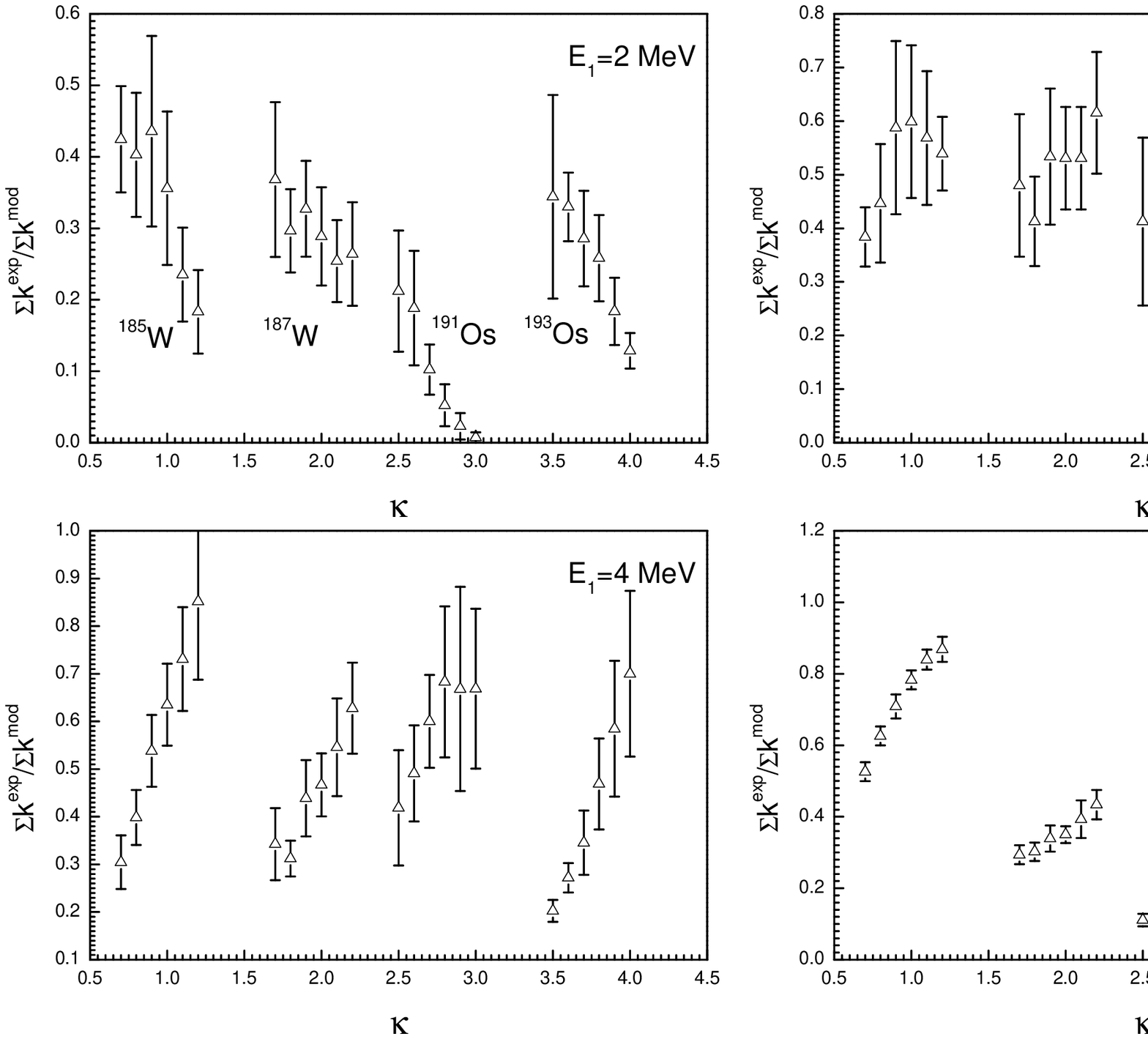}
\end{center}
\vspace{-5cm}
 Fig.~10.~The ratio between the experimental and calculated within model [11]
 and [12] radiative strength functions.
\end{figure}
\end{document}